\documentclass[aps,pra,twocolumn,showpacs,superscriptaddress,tightenlines,preprintnumbers,sort&compass,amsmath,amssymb,floatfix,nofootinbib]{revtex4}
\usepackage{graphicx}
\usepackage{color}

\def\extra#1{{``#1''}}

\def \etal{\textit{et al.}}

 \def\fig#1{{#1}}

\newcommand{\job}{J. Opt. B: Quantum Semiclass. Opt.~}
\newcommand{\sr}{Sci. Rep.}

\def\tr{{\rm Tr}}

\def\<{\langle}
\def\>{\rangle}
\def\vac{{\rm vac}}

\newcommand{\ket}[1]{\mbox{$|#1\rangle$}}
\newcommand{\bra}[1]{\mbox{$\langle#1|$}}
\newcommand\fra[2]{{\textstyle{\frac{#1}{#2}}}}

\def \info#1{}

\begin{document}

%\end{document}

%------------------------------------------------------------------

\title{Optimal two-qubit tomography based on local and global
measurements:\\
Maximal robustness against errors as described by condition
numbers}

\author{Adam Miranowicz}
\affiliation{CEMS, RIKEN, 351-0198 Wako-shi, Japan}
\affiliation{Faculty of Physics, Adam Mickiewicz University,
61-614 Pozna\'n, Poland}

\author{Karol Bartkiewicz}
\affiliation{Faculty of Physics, Adam Mickiewicz University,
61-614 Pozna\'n, Poland} \affiliation{RCPTM, Joint Laboratory of
Optics of Palack\'y University and Institute of Physics of AS CR,
Faculty of Science, Palack\'y University, 17. listopadu 12, 77-146
Olomouc, Czech Republic}

\author{Jan Pe\v{r}ina Jr.}
\affiliation{RCPTM, Joint Laboratory of Optics of Palack\'y
University and Institute of Physics of AS CR, Faculty of Science,
Palack\'y University, 17. listopadu 12, 77-146 Olomouc, Czech
Republic}

\author{Masato Koashi}
\affiliation{Photon Science Center, The University of Tokyo,
Bunkyo-ku, Tokyo 113-8656, Japan}

\author{Nobuyuki Imoto}
\affiliation{Graduate School of Engineering Science, Osaka
University, Toyonaka, Osaka 560-8531, Japan}

\author{Franco Nori}
\affiliation{CEMS, RIKEN, 351-0198 Wako-shi, Japan}
\affiliation{Department of Physics, The University of Michigan,
Ann Arbor, MI 48109-1040, USA}

\begin{abstract}
We present an error analysis of various tomographic protocols
based on the linear inversion for the reconstruction of an
unknown two-qubit state. We solve the problem of finding a
tomographic protocol which is the most robust against errors
in terms of the lowest value (i.e., equal to 1) of a
condition number, as required by the Gastinel-Kahan theorem.
In contrast, standard tomographic protocols, including those
based on mutually unbiased bases, are nonoptimal for
determining all 16 elements of an unknown two-qubit density
matrix. Our method is based on the measurements of the 16
generalized Pauli operators, where twelve of them can be
locally measured, and the other four require nonlocal Bell
measurements. Our method corresponds to selectively
measuring, one by one, all of the real and imaginary elements
of an unknown two-qubit density matrix. We describe two
experimentally feasible setups of this protocol for the
optimal reconstruction of two photons in an unknown
polarization state using conventional detectors and
linear-optical elements. Moreover, we define the operators
for the optimal reconstruction of the states of multiqubit or
multilevel (qudit) systems.

\end{abstract}

%\vspace{5mm}

\today


% 03.65.Wj  - Quantum tomography
% 03.67.-a  - Quantum information
% 42.50.Ex  - optical implementations

\maketitle

%------------------------------------------------------------------
\section{Introduction}

Quantum state tomography (QST) is a method of determining an
unknown quantum state (density matrix) in a series of
measurements on multiple copies of the state. QST is an
essential tool for the verification and benchmarking of
quantum devices used, e.g., for quantum state engineering,
quantum communication and quantum information processing.
Reviews on QST include Refs.~\cite{NielsenBook, ParisBook,
DAriano03}, and more recent results can be found in, e.g.,
Refs.~\cite{DAriano07, Bisio09,Cramer10, Gross10, Rehacek10,
Toth10,Liu12,Christandl12, Lundeen12, Salvail13} and
references therein. Dozens, if not hundreds, of QST protocols
have been proposed using various methods both for finite- and
infinite-dimensional quantum systems. Among photonic QST
protocols, those directly applicable to polarization qubits
have attracted considerable interest (for a review see
Ref.~\cite{Altepeter05review}). Here we analyze mainly
two-qubit tomography protocols and describe only their
photonic implementations with polarization qubits.

Applied QST is usually based on linear
inversion~\cite{NielsenBook} and maximum-likelihood
estimation~\cite{James01,Rehacek01,Blume10a, Teo11,Teo12,
Smolin12, Halenkova2012, Halenkova2012a, PerinaJr2012,
Bart12,Bart13a,Bart14}. Other proposals of QST are based on,
e.g., least-squares inversion~\cite{Opatrny97, PerinaJr2013a}
(which is also applied in the standard linear-inversion
approach to overdetermined systems), as well as Bayesian mean
estimation~\cite{ParisBook, Blume10b,Huszar12}, or linear
regression estimation~\cite{Qi13}.

Given this abundance of QST protocols, a natural question would be
which of them are optimal, according to some requirements or
criteria. The problem of the optimality of QST was studied from
different perspectives (see,
e.g.,~\cite{DAriano07,Bisio09,Roy07,Toth10,Blume10b,Klimov13})
including choosing optimal measurement sets to increase the
accuracy and efficiency of
estimation~\cite{Wootters89,Rehacek04,DAriano05,Ling06,Burgh08,Adamson10,Nunn10,Filippov10}.
Various quantitative approaches, in addition to the above
references, to testing the performances of QST protocols were
recently described by Bogdanov
\etal~\cite{Bogdanov10a,Bogdanov10b,Bogdanov11}.

In this paper we address the question of finding a QST
method, based on linear inversion, which is the most robust
against errors, as described by the condition number
$\kappa(A)$, later defined in Eq.~(\ref{kappa}) via the
spectral norm (i.e., the two-norm condition number). This QST
approach is based on solving a linear-system problem,
\begin{eqnarray}
  Ax = b,
\label{N1}
\end{eqnarray}
where $A$ is called here the \emph{rotation matrix} but is
also referred to as the \emph{coefficient matrix} or
\emph{data matrix} in more mathematical contexts. Moreover,
${b}$ is the \emph{observation vector}, which contains the
measured data, and $x={\rm vec}(\rho)$ is a real vector
describing the unknown state $\rho$ to be reconstructed.

Condition numbers are standard parameters characterizing the error
stability of, e.g., numerical
algorithms~\cite{AtkinsonBook,HighamBook,GolubBook}, which, in
particular, can be applied to QST based on linear
inversion~\cite{Bogdanov10a,Filippov10}. The significance of
applying a condition number in the error analysis of linear
systems explains the Gastinel-Kahan theorem~\cite{Kahan}, which
states that the relative distance of a nonsingular square matrix
$A$ to the set of singular matrices is given by the reciprocal of
the condition number. Thus, in particular, the condition number
$\kappa(A)$ is a measure of the QST robustness to errors in the
observation vector $b$. The smaller is the condition number the
more robust is the QST method, and the optimal method is described
by $\kappa(A)=1$.

To show the importance of error analysis in solving linear systems
$Ax=b$, let us analyze the following simple example:
\begin{eqnarray}
A&=&\left[
\begin{array}{cc}
 6 & 7 \\
 5 & 6 \\
\end{array}
\right]\quad \Rightarrow \quad A^{-1}=\left[
\begin{array}{cc}
 6 & -7 \\
 -5 & 6 \\
\end{array}
\right].
 \label{example1}
\end{eqnarray}
Then for two slightly different observation vectors $b$ one finds
two distinct solutions:
\begin{eqnarray*}
  b=\left[
\begin{array}{c}
 .7 \\
 .6 \\
\end{array}
\right] \Rightarrow
 x=\left[
\begin{array}{c}
 0 \\
 .1 \\
\end{array}
\right]
\; \& \;
b=\left[
\begin{array}{c}
 .71 \\
 .59 \\
\end{array}
\right] \Rightarrow
 x=\left[
\begin{array}{c}
 .13 \\
 -.01 \\
\end{array}
\right].
\end{eqnarray*}
It is clearly seen that these small relative changes in the
observation vector $b$ are amplified by one order in the
solution $x$. This unstable solution is a result of an
ill-conditioned linear system, as revealed by large condition
numbers (as defined below). For example, the condition number
based on the spectral norm is equal to $\kappa(A)\approx
146$. In general, the solution of a linear system $Ax=b$ is
most stable against changes (errors) in $b$ if a condition
number [say $\kappa(A)$] is equal to~1.

In this paper we describe an optimal two-qubit QST (referred
to as Protocol~1) based on both local and global measurements
to determine the mean values of some properly chosen
generalized Pauli operators (GPOs). These operators, except
four diagonal ones, correspond to the Gell-Mann operators for
the special unitary group SU(4) (see, e.g.,
Ref.~\cite{Kimura03}). The optimality of Protocol~1 refers to
the optimal value (i.e., equal to 1) of the condition number
based on the spectral norm.

In Table~I, we compare Protocol~1 with six other QST
protocols. In particular, we studied a QST method (referred
to here as Protocol~5) based on mutually unbiased bases
(MUB)~\cite{Wootters89,Bandyopadhyay01,Aravind03,DAriano05,Adamson10},
which requires both local and global measurements as
Protocol~1. Surprisingly, Protocol~5 is five times more
sensitive to errors than Protocol~1, and two-and-a-half times
worse than the QST based on the local measurements of tensor
products of the standard Pauli operators (referred to here as
Protocol~2).

Furthermore, we describe two feasible experimental setups for
performing Protocol~1 for the optimal reconstruction of an unknown
polarization state of two photons.

This paper is organized as follows: In Sec.~II, the QST
method based on a linear inversion is recalled. Section~III
introduces the concept of error analysis based on condition
numbers for QST. In Sec.~IV, the optimal nonlocal tomography
based on the measurements of GPOs is proposed as Protocol~1.
In Sec.~V, two setups of photonic implementations of
Protocol~1 are described. In Sec.~VI, we show how to
construct the operators for the optimal QST of the states of
multiqubit and multilevel (qudit) systems. A comparison of
Protocol~1 with some other QST methods is presented in the
concluding Sec.~VII. In Appendix~A, the GPOs and projectors
of all the discussed two-qubit protocols are summarized.
Beam-splitter transformations of entangled projectors are
described in Appendix~B.

\begin{widetext}
%------------------------------------------------------------------
\begin{table} % TABLE 1
\caption{Comparison of error robustness for various two-qubit QST
protocols.}
\begin{tabular}{c c c c c c c c}
  \hline  \hline
      Protocol \hspace{0mm} &
      Based on  &
      Projectors &
      Number of  \hspace{0mm} &
      local/global   \hspace{0mm} &
      condition no. &
      min[svd($C$)]&
      Eqs.
       \\
       &
       &
      &
       projectors &
      projectors&
      $\kappa(C)=\kappa^{2}(A)$ &
       &

       \\
\hline
   1 & optimal GPOs & $\gamma_k$ & 16 & local \& global & 1 & 1 & (\ref{protocol1sep}), (\ref{protocol1ent})
\\
   2 & Pauli operators & $\sigma_k\otimes\sigma_l$ & 16 & local & 2 & 1 & (\ref{protocol2})
\\
   3 & James \etal~basis~\cite{James01} & $\ket{\psi_k^{(3)}}$ & 16 & local & 60.1 & 0.1 & (\ref{protocol3})
\\
   4 & standard separable basis~\cite{Altepeter05,Burgh08} & $\ket{\psi_k^{(4)}}$ & 36 & local& 9&1 & (\ref{protocol4})
\\
   5 & mutually unbiased bases~\cite{Adamson10,Bandyopadhyay01}& $\ket{\psi_k^{(5)}}$ & 20 & local \& global&  5&1&  (\ref{protocol5})
   \\
   6 & Gell-Mann GPOs & $\Gamma^{(6)}_k$ & 16 & local \& global& 2&$\frac12$ & (\ref{protocol7})
\\
   7 & Patera-Zassenhaus GPOs & $\Gamma^{(7)}_k$ & 16 & local \& global& 2&4 & (\ref{protocol8})
\\
 \hline  \hline
\end{tabular}
\end{table}

%\twocolumngrid
\end{widetext}

%------------------------------------------------------------------
\section{Principles of QST based on linear inversion}

The numerical procedure to reconstruct a density matrix $\rho$
from experimental data has been widely used in quantum state
engineering (see, e.g., Ref.~\cite{ParisBook} and references
therein).

First, it is useful to represent $\rho$ as a vector. This
procedure can be seen as representing an operator $\rho$ in
Hilbert space as a superoperator $x= {\rm vec}(\rho)$ in
Liouville space. The matrix-to-vector operation, which we
denote by ${\rm vec}(\rho)$, can be given by an arbitrarily
chosen reordering of the elements $\rho_{ij}$ of $\rho$. For
example, one can choose the standard order of $\rho_{ij}$,
i.e., $x'={\rm vec}'(\rho) =
[\rho_{00},\rho_{01},\ldots,\rho_{32},\rho_{33}]^T$ for a
two-qubit state $\rho$. The above vector is complex and
contains redundant information as $\rho_{ij}=\rho_{ji}^*$.
Thus, it is convenient to transform $\rho$ into a real
vector, e.g., as follows:
\begin{eqnarray}
x={\rm vec}(\rho) = [\rho_{00},{\rm Re} \rho_{01},{\rm Im}
\rho_{01},{\rm Re} \rho_{02},{\rm Im} \rho_{02},
\ldots,\rho_{33}]^T, \label{Na1}
\end{eqnarray}
where only the elements $\rho_{ij}$ for $i\le j$ are included.
Obviously, any other ordering can be applied but it should be used
consistently.

To find all the elements of ${x}\equiv[x_i]_{16 \times 1}$, one
has to solve the set of linear equations, given in Eq.~(\ref{N1}),
where now the rotation matrix is ${A}\equiv[A_{j i}]_{N_{\rm
eqs}\times 16}$ and the observation vector is
${b}\equiv[b_{j}]_{N_{\rm eqs} \times 1}$. Specifically, the
element $A_{j i}$ is the coefficient of $x_i$ in the $j$th
equation for a chosen tomographic rotation, while a given element
$b_j$ of the observation vector $b$ can correspond, e.g., to
coincidence photocounts in optical experiments or the integrated
area of spectra in the spectroscopy of nuclear magnetic resonance
(NMR). Let us assume that there are $N_{\rm r}$ readouts and each
of them yields $N_{\rm vals}$ values, which can correspond to,
e.g., coincidence counts in photon detectors or the number of
peaks in the real and imaginary parts of an NMR spectrum. Then the
number of equations, $N_{\rm eqs}$, is equal to $N_{\rm r}\times
N_{\rm vals}$. Formally, extra equations can be added, which
correspond to the normalization condition, $\tr\rho=1$.

This problem is usually \emph{overdetermined} if there are more
equations than unknowns. The redundant expressions can (sometimes)
enable more accurate reconstruction of $x$. By applying standard
least-squares-fitting analysis one obtains
\begin{eqnarray}
  {C} x=\tilde b \quad{\rm with}\quad {C} ={A}^\dagger {A},\quad \tilde b
  ={A}^\dagger {b},
\label{N09}
\end{eqnarray}
where the overdeterminacy is removed as $\tilde{b}\equiv[\tilde
b_{j}]_{16 \times 1}$ and ${C}\equiv[ {C}_{ij}]_{16 \times 16}$.
The matrix $C$ is sometimes referred to as the \emph{error
matrix}~\cite{Long01}. Thus, to reconstruct a density matrix
$\rho$, it is enough to calculate
\begin{eqnarray}
  x = {C}^{-1}\tilde  b \;\;\rightarrow\;\; \rho={\rm vec}^{-1}(x),
\label{N10}
\end{eqnarray}
where ${\rm vec}^{-1}(x)$ is the operation inverse to ${\rm
vec}(x)$. The least-squares analysis is based on the
minimalization of $\chi^{2}=||Ax-b||^2$.

%------------------------------------------------------------------
\section{Error analysis of QST based on linear inversion}

Here let us address the question of how the experimental errors
are magnified through the numerical procedure of linear inversion.
Thus, the problem now is about the reliability of the
reconstructed density matrix $\rho$ corresponding to the vector
$x=A^{-1}b$ for a given set of rotations $A$ (representing our
linear tomographic system) and for the measured data $b$.

Even a simple application of a singular-value decomposition of a
nonsingular square matrix $A\in {\cal R}^{n\times n}$,
\begin{eqnarray}
A=UD V^T=\sum_{i=1}^n u_i\bar\sigma_i v^T_i,
\label{svd1}
\end{eqnarray}
implies that the error robustness can be related to the minimal
singular value, $\min_i(\bar\sigma_i)\equiv\sigma_{\min}(A)$. This
can be seen by the expansion~of the solution $x$~\cite{GolubBook}:
\begin{eqnarray}
x=A^{-1}b=(V D^{-1} U^T)b=\sum_{i=1}^n
\frac{u^{T}_ib}{\bar\sigma_i}v_i.
\label{svd2}
\end{eqnarray}
Here, $U=[u_1,\ldots,u_n]$ and $V=[v_1,\ldots,v_n]$ are the
left- and right-hand singular vectors for $A$, respectively,
and $D={\rm diag}([\bar\sigma_1,\ldots,\bar\sigma_n])$ is a
diagonal matrix of the singular values $\bar\sigma_i$ for $A$
(which should not be confused with $\sigma_i$ denoting the
Pauli operators). Thus, by assuming $\sigma_{\min}\ll 1$,
small errors in $A$ or $b$ can induce relatively large errors
in $x$.

As an indicator of the error robustness (or error sensitivity) of
QST methods we apply the \emph{condition number}, which is defined
for a nonsingular square matrix $A$ as
follows~\cite{AtkinsonBook,HighamBook,GolubBook}:
\begin{equation}
  {\rm cond}_{\alpha,\beta}(A) = \Vert A \Vert_{\alpha,\beta}\; \Vert A^{-1}
  \Vert_{\beta,\alpha},
\label{N11}
\end{equation}
where the convention is used that ${\rm cond}_{\alpha,\beta}(A) =
+\infty$ for a singular matrix $A$. The subordinate matrix norm
$\Vert \cdot \Vert_{\alpha,\beta}$ in Eq.~(\ref{N11}) can be given
by the vector norms:
\begin{equation}
  \Vert A \Vert_{\alpha,\beta} = \max_{x\neq 0} \frac{\Vert Ax \Vert_{\beta}}{\Vert x
  \Vert_{\alpha}}.
\label{N12}
\end{equation}
The condition number was introduced by Turing~\cite{Turing} for
the Frobenius norm, but it clearly depends on the underlying norm.
Note that the property
\begin{equation}
{\rm cond}_{\alpha,\beta}(A)\ge 1
\end{equation}
holds for any norm. The condition number has an algebraic
interpretation as a normalized Fr\'echet derivative of the map $A
\rightarrow A^{-1}$~\cite{GolubBook}.

The importance of ${\rm cond}_{\alpha,\beta}$ in the linear-system
error-robustness analysis is based on the Gastinel-Kahan
theorem~\cite{Kahan} (see also
Refs.~\cite{AtkinsonBook,HighamBook,GolubBook}), which states that
the relative distance of a nonsingular square matrix $A$ to the
set of singular matrices,
\begin{equation}
  {\rm dist}_{\alpha,\beta}(A) := \min \left\{
  \frac{\Vert A -P\Vert_{\alpha,\beta}}
 {\Vert A \Vert_{\alpha,\beta}}: P \;{\rm is\;
 singular}\right\},
\label{N13}
\end{equation}
is the reciprocal of the condition number,
\begin{eqnarray}
  {\rm dist}_{\alpha,\beta}(A) &=&
  \frac{1}{{\rm cond}_{\alpha,\beta}(A)}.
\label{N14}
\end{eqnarray}
In our physical context, the condition number can roughly be
interpreted as the rate at which the reconstructed density
matrix $x$ in a given QST method $Ax=b$ changes with a change
in the observation vector $b$. If ${\rm
cond}_{\alpha,\beta}(A)$ is small, the QST method (or the
corresponding rotation matrix $A$) is called
\emph{well-conditioned}, which implies that the system is
robust against errors in the observation vector $b$. However,
the problem is referred to as \emph{ill-conditioned} if ${\rm
cond}_{\alpha,\beta}(A)$ is large, and \emph{ill-posed} if
${\rm cond}_{\alpha,\beta}(A)$ is infinite. In contrast to a
well-conditioned method, an ill-conditioned QST method has a
solution $x$ sensitive to errors (or unstable) in $b$, so
even a small error in $b$ can cause a large error in $x$.

To show the operational (or physical) importance of condition
numbers more explicitly, let us recall a well known theorem
(Theorem 8.4 in Ref.~\cite{AtkinsonBook}): Consider the system
given in Eq.~(\ref{N1}) with nonsingular $A$. Assume perturbations
$\delta\,b$ in $b$ and $\delta\, A$ in $A$, such that $||\delta\,
A||<1/||A^{-1}||$ implying that $A+\delta\, A$ is nonsingular. If
perturbations $\delta\, x$ are defined implicitly by
\begin{eqnarray}
  (A+\delta\, A)(x+\delta\, x) &=& b+\delta\, b,
\label{Atkinson1}
\end{eqnarray}
then
\begin{equation}
 \frac{||\delta\, x||}{||x||} \le \frac{{\rm cond}_{\alpha,\beta}(A)}
 {1-{\rm cond}_{\alpha,\beta}(A)\frac{||\delta\, A||}{||A||}}
  \left(\frac{||\delta\, A||}{||A||}+\frac{||\delta\, b||}{||b||} \right).
\label{Atkinson2}
\end{equation}
By ignoring perturbations in $A$, the lower and upper bounds for
the relative perturbations in $x$ are simply given
by~\cite{AtkinsonBook}:
\begin{equation}
 \frac{1}{{\rm cond_{\alpha,\beta}}(A)}
 \frac{||\delta\, b||}{||b||} \le  \frac{||\delta\, x||}{||x||} \le {\rm cond_{\alpha,\beta}}(A)
 \frac{||\delta\, b||}{||b||},
\label{Atkinson3}
\end{equation}
where the right-hand inequality is a special case of the
inequality in Eq.~(\ref{Atkinson2}). Thus, if a condition number
${\rm cond_{\alpha,\beta}}(A)$ is equal (or very close) to one,
then small relative changes in the observation vector $b$ imply
equally small relative changes in the reconstructed state $x$.

Below, we apply the spectral norm (also called the two-norm)
given by the largest singular value of $A$, i.e,
\begin{equation} \Vert A \Vert_{2,2}\equiv \Vert A
\Vert_{2}=\max[{\rm svd}(A)]\equiv \sigma_{\max}(A),
\end{equation}
where the function ${\rm svd}(A)$ returns the singular values of
$A$. Then the condition number ${\rm cond}_{2,2}(A)\equiv {\rm
cond}_2(A)$ can be given by a simple formula
\begin{eqnarray}
  \kappa(A) \equiv {\rm cond}_2(A) &=& \frac{\sigma_{\max}(A)}{\sigma_{\min}(A)},
\label{kappa}
\end{eqnarray}
which is a special case of Eq.~(\ref{N11}) because
\begin{equation}
\Vert A^{-1} \Vert_{2}= \max[{\rm svd}(A^{-1})]=\frac{1}{\min[{\rm
svd}(A)]}\equiv \frac{1}{\sigma_{\min}(A)}.
\end{equation}
Note that Eq.~(\ref{kappa}) can be applied not only to square
matrices but also to nonsquare ones; e.g., to the rotation
matrices $A$ of the dimensions $36\times 16$ and $20\times
16$ for Protocols~4 and~5, respectively, as listed in
Table~I.

Singular values reveal some important aspects of the geometry of a
linear transformation $A$. In particular, $\kappa(A)$, given in
Eq.~(\ref{kappa}) for a square matrix $A$, has a clear geometrical
interpretation as a degree of the distortion of a unit sphere (or
rather hyper-sphere) under the transformation by
$A$~\cite{MeyerBook}; or, equivalently, as a measure of the
elongation of the hyper-ellipsoid $\{Ax: \Vert x
\Vert_{2}=1\}$~\cite{GolubBook}.

One could also calculate the condition number defined via other
norms, e.g., ${\rm cond}_F(A) = \Vert A \Vert_F\; \Vert A^{-1}
\Vert_F$, based on the Frobenius norm $\Vert A \Vert_{F}^2=\sum_i
\bar\sigma_i^2$. However, for brevity, we apply in this paper only
the condition number $\kappa$, defined in Eq.~(\ref{kappa}).

As explained above, the smallest eigenvalue of ${C}$ (or $A$) can
also be considered an ``error robustness parameter''.  One can
write this parameter  as the smallest singular value of ${C}$:
\begin{eqnarray}
  \sigma_{\min}({C}) = \min[{\rm svd}({C})] = ||{C}^{-1}||_2.
\label{N16}
\end{eqnarray}
The condition numbers, in contrast to Eq.~(\ref{N16}), also
contain information about the range of the eigenvalues of
${C}$. In the context of tomographic reconstructions, the
parameter $\sigma_{\min}({C})$ was applied in, e.g.,
Refs.~\cite{Wildenthal88,Long01}.

One can raise the question of whether a condition number
${\rm cond_{\alpha,\beta}}(A)$ or the minimum singular value
$\sigma_{\min}(A)$ is more appropriate in the analysis of
errors in linear systems. Some justifications, like those in
Eq.~(\ref{svd2}) and below Eq.~(\ref{N14}), are applicable to
both ${\rm cond_{\alpha,\beta}}(A)$ and $\sigma_{\min}(A)$.
However, inequalities in Eqs.~(\ref{Atkinson2})
and~(\ref{Atkinson3}) clearly show the advantage of using
${\rm cond_{\alpha,\beta}}(A)$ over $\sigma_{\min}(A)$. Yet
another simple argument in support of ${\rm
cond_{\alpha,\beta}}(A)$ can be given as follows: Let us
rescale vector $b$ to be ten times its original values. Then
$A$ is also enlarged by 10. This changes $\sigma_{\min}(A)$,
but the condition number ${\rm cond_{\alpha,\beta}}(A)$
remains unchanged.
%------------------------------------------------------------------
\begin{table}  % TABLE 2
\caption{How to project a given state $\rho$ of two
polarization qubits onto all the \emph{separable} eigenstates
$\ket{\psi_{kl}}$ of the optimal GPOs $\gamma_k$
($k=1,\ldots,12$) in the implementation of Protocol~1 shown
in Figs.~1 (Setup~1) and~2 (Setup~2): Rotate locally $\rho$
by the angles specified below for the HWPs ($H_1$ and $H_2$)
and the QWPs ($Q_1$ and $Q_2$); and then project the rotated
state onto $\ket{00}\equiv\ket{HH}$. This probabilistic
projection occurs when both detectors $D_{1H}$ and $D_{2H}$
($D_{1}$ and $D_{2}$) click in Setup~1 (Setup~2).}
\begin{tabular}{c | c | c c | c c}
  \hline  \hline
      local &
      eigenstates $\ket{\psi_{kl}}$   &
      \multicolumn{2}{|c|}{qubit 1} &
      \multicolumn{2}{|c}{qubit 2}
       \\
      optimal GPOs &
      of optimal GPOs  &
      $H_1$ &
      $Q_1$ &
      $H_2$ &
      $Q_2$
       \\
\hline
      $\gamma_1$ &
      $\ket{00}$  &
      0 &
      0 &
      0 &
      0
\\
      $\gamma_2$ &
      $\ket{01}$  &
      0 &
      0 &
      45$^0$ &
      0
\\
      $\gamma_3$ &
      $\ket{10}$  &
      45$^0$ &
      0 &
      0 &
      0
\\
      $\gamma_4$ &
      $\ket{11}$  &
      45$^0$ &
      0 &
      45$^0$ &
      0
\\
      $\gamma_5$ &
      $\ket{0+}$  &
      0 &
      0 &
      22.5$^0$ &
      0
\\
       &
      $\ket{0-}$  &
      0 &
      0 &
      67.5$^0$ &
      0
\\
      $\gamma_6$ &
      $\ket{0R}$  &
      0 &
      0 &
      0 &
      45$^0$
\\
       &
      $\ket{0L}$  &
      0 &
      0 &
      0 &
      -45$^0$
\\
      $\gamma_7=\gamma'_{15}$ &
      $\ket{\!+\!0}$  &
      22.5$^0$ &
      0 &
      0 &
      0
\\
       &
      $\ket{\!-\!0}$  &
      67.5$^0$ &
      0 &
      0 &
      0
\\
      $\gamma_8=\gamma'_{16}$ &
      $\ket{R0}$  &
      0 &
      45$^0$ &
      0 &
      0
\\
       &
      $\ket{L0}$  &
      0 &
      -45$^0$ &
      0 &
      0
\\
      $\gamma_9$ &
      $\ket{1+}$  &
      45$^0$ &
      0 &
      22.5$^0$ &
      0
\\
       &
      $\ket{1-}$  &
      45$^0$ &
      0 &
      67.5$^0$ &
      0
\\
      $\gamma_{10}$ &
      $\ket{1R}$  &
      45$^0$ &
      0 &
      0 &
      45$^0$
\\
       &
      $\ket{1L}$  &
      45$^0$ &
      0 &
      0 &
      -45$^0$
\\
      $\gamma_{11}=\gamma'_{13}$ &
      $\ket{+1}$  &
      22.5$^0$ &
      0 &
      45$^0$ &
      0
\\
       &
      $\ket{-1}$  &
      67.5$^0$ &
      0 &
      45$^0$ &
      0
\\
      $\gamma_{12}=\gamma'_{14}$ &
      $\ket{R1}$  &
      0 &
      45$^0$ &
      45$^0$ &
      0
\\
       &
      $\ket{L1}$  &
      0 &
      -45$^0$ &
      45$^0$ &
      0
%------------------------------------------------------------------
\\
 \hline  \hline
\end{tabular}
\end{table}
%------------------------------------------------------------------
\begin{table}  % TABLE 3
\caption{ How to project $\rho$ onto all the \emph{entangled}
eigenstates $\ket{\psi_{kl}}$ of the optimal GPOs $\gamma_k$
($k=13,\ldots,16$) in Setup~1: Rotate locally $\rho$ by the
angles specified below for the HWPs and QWPs, and then
project the rotated state onto the singlet state
$\ket{\Psi^{-}}$. The desired projection is heralded by
single clicks in both detectors $D_{1H}$ and $D_{2V}$ or
$D_{1V}$ and $D_{2H}$. }
\begin{tabular}{c | c | c c | c c}
  \hline  \hline
      nonlocal &
      eigenstates $\ket{\psi_{kl}}$   &
      \multicolumn{2}{|c|}{qubit 1} &
      \multicolumn{2}{|c}{qubit 2}
       \\
      optimal GPOs &
      of optimal GPOs  &
      $H^{}_1$ &
      $Q^{}_1$ &
      $H^{}_2$ &
      $Q^{}_2$
       \\
\hline
      $\gamma_{13}$ &
      $\ket{\Psi^{-}}$  &
      0 &
      0 &
      0 &
      0
\\

       &
      $\ket{\Psi^{+}}$  &
      45$^0$ &
      -45$^0$ &
      $0$ &
      45$^0$

\\
      $\gamma_{14}$ &
      $\ket{\bar\Psi^{-}}$  &
      0 &
      45$^0$ &
      -22.5$^0$ &
      $0$
\\
       &
      $\ket{\bar\Psi^{+}}$  &
      0 &
      45$^0$ &
      22.5$^0$ &
      90$^0$
\\
      $\gamma_{15}$ &
      $\ket{\Phi^{-}}$  &
      0 &
      -45$^0$ &
      0 &
      45$^0$
\\
       &
      $\ket{\Phi^{+}}$  &
      45$^0$ &
      0 &
      0 &
      0
\\
      $\gamma_{16}$ &
      $\ket{\bar\Phi^{-}}$  &
      0 &
      45$^0$ &
      -22.5$^0$ &
      90$^0$
\\
       &
      $\ket{\bar\Phi^{+}}$  &
      0 &
      45$^0$ &
      22.5$^0$ &
      0
%------------------------------------------------------------------
\\
 \hline  \hline
\end{tabular}
\end{table}

%------------------------------------------------------------------
\section{Optimal nonlocal tomography}

Now, let us describe the main result of this paper, i.e., a
proposal of an optimal two-qubit QST (referred to as
Protocol~1), which is maximally robust against errors, as
described by the condition number $\kappa(A)$ equal to 1.

This protocol requires both local and nonlocal measurements
corresponding to measuring the following generalized Pauli
operators (GPOs). There are twelve separable (local) GPOs:
\begin{eqnarray}
\gamma_{1} & = & |00\rangle\langle00|,\quad\quad\quad\gamma_{2}=|01\rangle\langle01|,
 \nonumber \\
\gamma_{3}  & = &  |10\rangle\langle10|,\quad\quad\quad\gamma_{4}=|11\rangle\langle11|,
 \nonumber \\
\gamma_{5} & = & \frac{1}{2}|0\rangle\langle0|\otimes\sigma_{1},\quad
\gamma_{6}  =  \frac{1}{2}|0\rangle\langle0|\otimes\sigma_{2},\nonumber \\
\gamma_{7} & = & \frac{1}{2}\sigma_{1}\otimes|0\rangle\langle0|,\quad
\gamma_{8}  =  \frac{1}{2}\sigma_{2}\otimes|0\rangle\langle0|,\nonumber \\
\gamma_{9} & = & \frac{1}{2}|1\rangle\langle1|\otimes\sigma_{1},\quad
\gamma_{10}  =  \frac{1}{2}|1\rangle\langle1|\otimes\sigma_{2},\nonumber \\
\gamma_{11} & = & \frac{1}{2}\sigma_{1}\otimes|1\rangle\langle1|,\quad
\gamma_{12}  =  \frac{1}{2}\sigma_{2}\otimes|1\rangle\langle1|,
\label{protocol1sep}
\end{eqnarray}
and four entangled (global) operators
\begin{eqnarray}
\gamma_{13} & = & \tfrac12(|\Psi^{+}\rangle\langle\Psi^{+}|-|\Psi^{-}\rangle\langle\Psi^{-}|),\nonumber\\
\gamma_{14}&=&  \tfrac12(|\bar\Psi^{+}\rangle\langle\bar\Psi^{+}|
-|\bar\Psi^{-}\rangle\langle\bar\Psi^{-}|),\nonumber\\
\gamma_{15}
&=& \tfrac12(|\Phi^{+}\rangle\langle\Phi^{+}|-|\Phi^{-}\rangle\langle\Phi^{-}|),\nonumber\\
\gamma_{16} & = &
 \tfrac12(|\bar\Phi^{+}\rangle\langle\bar\Phi^{+}|
-|\bar\Phi^{-}\rangle\langle\bar\Phi^{-}|),\nonumber\\
\label{protocol1ent}
\end{eqnarray}
where $\sigma_n$ are the standard (single-qubit) Pauli operators,
$|\Phi^{\pm}\rangle=(|00\rangle\pm |11\rangle)/\sqrt{2}$ and
$|\Psi^{\pm}\rangle=(|01\rangle\pm |10\rangle)\sqrt{2}$ are the
Bell states, $|\bar\Phi^{\pm}\rangle=(S\otimes
I)|\Phi^{\pm}\rangle=(|00\rangle\pm i|11\rangle)\sqrt{2}$ and
$|\bar\Psi^{\pm}\rangle=(S\otimes I)|\Psi^{\pm}\rangle
=(|01\rangle\pm i|10\rangle)/\sqrt{2}$ are Bell-like states, which
are given in terms of the phase gate $S=\ket{0}\bra{0}+i
\ket{1}\bra{1}$. The set of 16 operators is Hermitian and
orthogonal in the Hilbert-Schmidt inner product, just as the set
$\{\sigma_1,\sigma_2,\sigma_3,I\}$. Thus, we refer to the former
set as GPOs, although it does not include the identity operator.
For clarity, all these 16 GPOs  are given explicitly in the
standard Fock basis in Appendix~A. In terms of the two-qubit
density matrix $\rho$, each $\gamma_{j}$ monitors either the real
or imaginary part of a matrix element $\rho_{kl}$ of the density
matrix written in a common basis.

According to the convention, given in Eq.~(\ref{Na1}), a
two-qubit density matrix $\rho$ can be represented as a real
vector $x=(x_1,\ldots,x_{16})$ with its elements given as
follows
\begin{eqnarray}
  \rho = \left[
\begin{array}{cccc}
 x_{1} & x_{2}+i x_{3} & x_{4}+i x_{5} & x_{6}+i x_{7} \\
 x_{2}-i x_{3} & x_{8} & x_{9}+i x_{10} & x_{11}+i x_{12} \\
 x_{4}-i x_{5} & x_{9}-i x_{10} & x_{13} & x_{14}+i x_{15} \\
 x_{6}-i x_{7} & x_{11}-i x_{12} & x_{14}-i x_{15} & x_{16} \\
\end{array}
\right]\!\!. \label{rho}
\end{eqnarray}
Then, the mean values $b_k=\tr(\rho\gamma_k)$ are simply related
to $x_l$ as
\begin{align}
&b_{1}=x_{1}  &&b_{2}=x_{8},&& b_{3}=x_{13},&& b_{4}=x_{16},
\nonumber \\
&b_{5}=x_{2}, &&b_{6}=-x_{3},&& b_{7}=x_{4},&& b_{8}=-x_{5},
\nonumber \\
&b_{9}=x_{14},&&b_{10}=-x_{15},&& b_{11}=x_{11},&& b_{12}=-x_{12},
\nonumber \\
&b_{13}=x_{9},&&b_{14}=-x_{10},&& b_{15}=x_{6},&& b_{16}=-x_{7}.
\label{b_vector}
\end{align}
It is seen that our method corresponds to selectively
measuring, one by one, all the real and imaginary elements of
an unknown two-qubit density matrix. Thus, solving this
linear problem $b=Ax$ is trivial because $A^{-1}=A^T$, where
\begin{eqnarray}
A=\left[
\begin{array}{cccccccccccccccc}
 1 & 0 & 0 & 0 & 0 & 0 & 0 & 0 & 0 & 0 & 0 & 0 & 0 & 0 & 0 & 0 \\
 0 & 0 & 0 & 0 & 0 & 0 & 0 & 1 & 0 & 0 & 0 & 0 & 0 & 0 & 0 & 0 \\
 0 & 0 & 0 & 0 & 0 & 0 & 0 & 0 & 0 & 0 & 0 & 0 & 1 & 0 & 0 & 0 \\
 0 & 0 & 0 & 0 & 0 & 0 & 0 & 0 & 0 & 0 & 0 & 0 & 0 & 0 & 0 & 1 \\
 0 & 1 & 0 & 0 & 0 & 0 & 0 & 0 & 0 & 0 & 0 & 0 & 0 & 0 & 0 & 0 \\
 0 & 0 & s & 0 & 0 & 0 & 0 & 0 & 0 & 0 & 0 & 0 & 0 & 0 & 0 & 0 \\
 0 & 0 & 0 & 1 & 0 & 0 & 0 & 0 & 0 & 0 & 0 & 0 & 0 & 0 & 0 & 0 \\
 0 & 0 & 0 & 0 & s & 0 & 0 & 0 & 0 & 0 & 0 & 0 & 0 & 0 & 0 & 0 \\
 0 & 0 & 0 & 0 & 0 & 0 & 0 & 0 & 0 & 0 & 0 & 0 & 0 & 1 & 0 & 0 \\
 0 & 0 & 0 & 0 & 0 & 0 & 0 & 0 & 0 & 0 & 0 & 0 & 0 & 0 & s & 0 \\
 0 & 0 & 0 & 0 & 0 & 0 & 0 & 0 & 0 & 0 & 1 & 0 & 0 & 0 & 0 & 0 \\
 0 & 0 & 0 & 0 & 0 & 0 & 0 & 0 & 0 & 0 & 0 & s & 0 & 0 & 0 & 0 \\
 0 & 0 & 0 & 0 & 0 & 0 & 0 & 0 & 1 & 0 & 0 & 0 & 0 & 0 & 0 & 0 \\
 0 & 0 & 0 & 0 & 0 & 0 & 0 & 0 & 0 & s & 0 & 0 & 0 & 0 & 0 & 0 \\
 0 & 0 & 0 & 0 & 0 & 1 & 0 & 0 & 0 & 0 & 0 & 0 & 0 & 0 & 0 & 0 \\
 0 & 0 & 0 & 0 & 0 & 0 & s & 0 & 0 & 0 & 0 & 0 & 0 & 0 & 0 & 0 \\
\end{array}
\right] \label{A_2qubits}
\end{eqnarray}
with $s=-1$. This implies that all the singular
values of $A$ are equal to 1, so the condition number is minimal,
$\kappa(A)=1$. For this reason, Protocol~1 is referred to as
optimal.

Note that all the nonlocal operators, given in
Eq.~(\ref{protocol1ent}), are related by local operations.  For
example, they can be expressed in terms of $\gamma_{13}$ as
follows:
\begin{eqnarray}
\gamma_{14}&=&(S\otimes I) \gamma_{13} (S^\dagger\otimes I),
  \nonumber\\
\gamma_{15}
&=&(I\otimes \sigma_1) \gamma_{13} (I\otimes \sigma_1),
 \nonumber\\
\gamma_{16} & = &
(S\otimes \sigma_1) \gamma_{13} (S^\dagger\otimes \sigma_1),
\end{eqnarray}
where $S=|0\rangle\langle0|+i |1\rangle\langle1|$ is the phase
gate, and $I$ is the single-qubit identity operator. More
importantly, they can be disentangled by applying the
controlled-NOT (CNOT) gate, $U_{\rm CNOT}$ and changed into some
local GPOs, given in Eq.~(\ref{protocol1sep}) as follows
\begin{eqnarray}
U_{\rm CNOT}\gamma_{k}U_{\rm CNOT}&=&\gamma_{k'},
\end{eqnarray}
where $(k,k')=(13,11)$, $(14,12)$,  $(15,7)$, and $(16,8)$.
All the 28 eigenstates (projectors) of the optimal GPOs are
listed in Tables~II and~III. In particular, those for
$\gamma_{5},\ldots,\gamma_{12}$ are given by
Eq.~(\ref{protocol1sep}) after applying the eigenstate
expansions of the standard Pauli operators
\begin{eqnarray}
  \sigma_1 &=& \ket{+}\bra{+}-\ket{-}\bra{-}=2\ket{+}\bra{+}-I=I-2\ket{-}\bra{-},
\nonumber \\
  \sigma_2 &=& \ket{L}\bra{L}-\ket{R}\bra{R}=2\ket{L}\bra{L}-I=I-2\ket{R}\bra{R},
\nonumber \\
  \sigma_3 &=& \ket{0}\bra{0}-\ket{1}\bra{1}=2\ket{0}\bra{0}-I=I-2\ket{1}\bra{1},
\label{pauli}
\end{eqnarray}
where $\ket{\pm}=(\ket{0}\pm\ket{1})/\sqrt{2}$,
$\ket{R}=(\ket{0}-i\ket{1})/\sqrt{2}$, and
$\ket{L}=(\ket{0}+i\ket{1})/\sqrt{2}$, which can be interpreted,
respectively, as diagonal, antidiagonal, right-circular, and
left-circular polarization states for the optical polarization
qubits.

We note that the number of 28 projectors can be reduced, e.g., by
applying the identity resolutions given in Eq.~(\ref{pauli}).
However, we use this complete set of eigenstates for the same
reason of improved experimental stability, as in the case of the
application of the standard separable QST (Protocol~4) based on
the projections onto all 36 tensor products of the eigenstates of
the standard single-qubit Pauli
operators~\cite{Altepeter05,Burgh08}.

%------------------------------------------------------------------
\begin{figure}

 \fig{ \includegraphics[width=8cm]{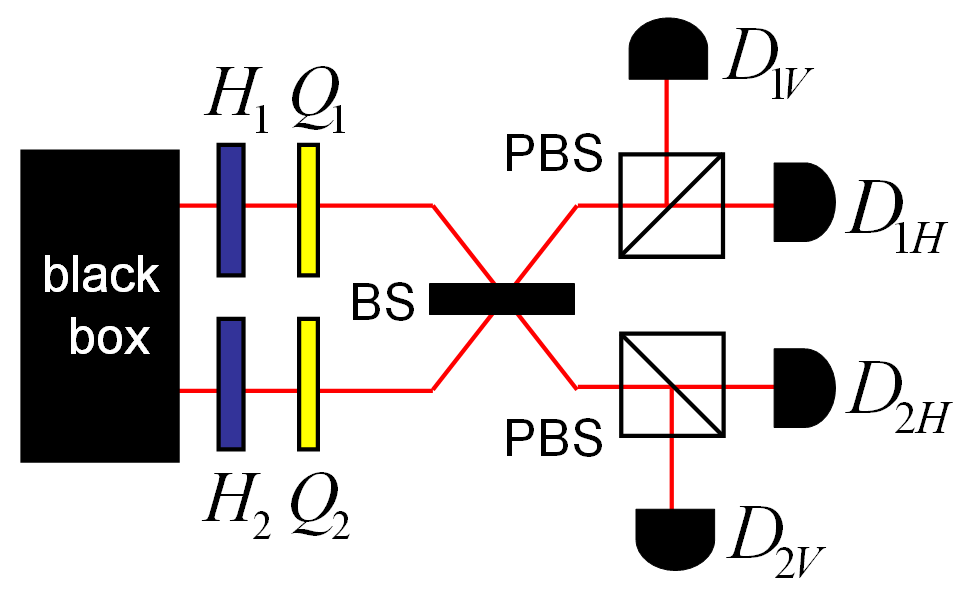}}

\caption{(Color online) Setup~1 for the experimental
implementation of the optimal QST (Protocol~1) of an unknown
two-photon polarization state $\rho$. Here the state $\rho$
is the output of a ``black box'' system, PBSs denote
polarizing beam splitters, $D_{1p}$ and $D_{2p}$ (with
$p=H,V$) correspond to detectors, whose outputs are connected
to a coincidence counter (for simplicity, not plotted here).
Moreover, $Q_1$ and $Q_2$ denote the quarter-wave plates
(QWPs) and $H_1$ and $H_2$ stand for the half-wave plates
(HWPs). A balanced ($50:50$) nonpolarizing beam splitter (BS)
is used for the Bell measurement of the nonlocal projectors
$\gamma_n$ for $n=13,\ldots,16$, given in
Eq.~(\ref{protocol1ent}). Here we assume that this BS is
removed if the local projectors $\gamma_n$ for
$n=1,\ldots,12$, as listed in Table~II, are measured. This
method formally corresponds to rotating $\rho$ by the HWPs
and QWPs at the angles specified in Tables~II and~III; and
then projecting them at the fixed states $\ket{\psi_{\rm
fixed}}=\ket{00}$ (if the BS is removed) and $\ket{\psi_{\rm
fixed}}=\ket{\Psi^{-}}$ (if the BS is inserted),
respectively. This approach formally corresponds to
projecting $\rho$ onto all the 28 eigenstates
$\ket{\psi_{kl}}$, from which the mean values of the 16
optimal GPOs, $\gamma_k$, can be directly calculated. The
projection onto the separable state $\ket{00}$ is heralded by
the coincidence clicks in the detectors $D_{1H}$ and
$D_{2H}$, while the projection onto the singlet state
$\ket{\Psi^{-}}$ occurs for the coincidence clicks in either
pair of the detectors: $D_{1H}$ and $D_{2V}$ or $D_{1V}$ and
$D_{2H}$.} \label{fig1}
\end{figure}
%------------------------------------------------------------------
\begin{figure}

 \fig{ \includegraphics[width=7cm]{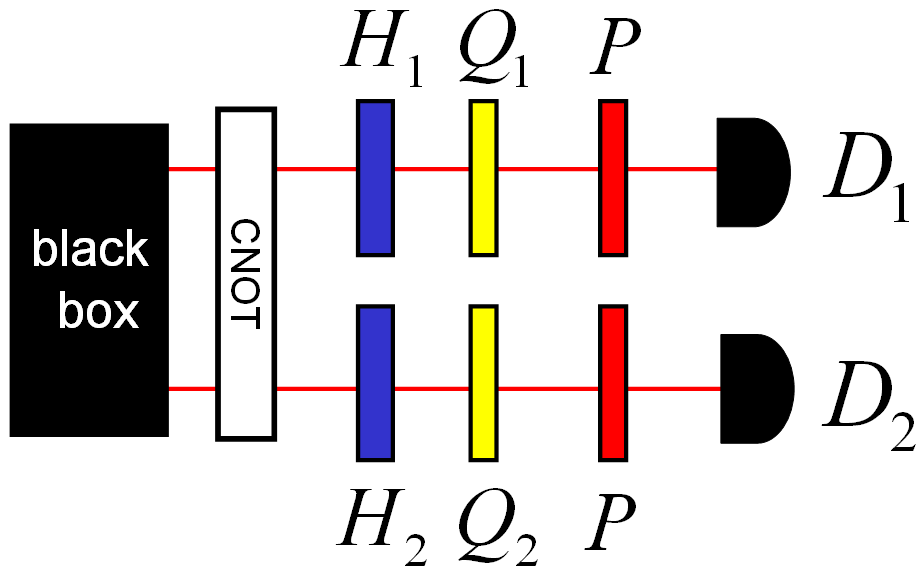}}

\caption{(Color online)  Setup~2 implementing Protocol~1,
analogous to Setup~1 in Fig.~1, but here all the rotated
states are projected onto the \emph{same} state via the
polarizers $P$ that transmit only photons of one polarization
(say, horizontal corresponding to $\ket{0}\equiv \ket{H}$).
The CNOT gate is used for disentangling the
maximally-entangled projectors $\ket{\psi_{kl}}$ for
$n=13,\ldots,16$, given in Eq.~(\ref{protocol1ent}). This
CNOT should be removed for measuring the local projectors
$\gamma_n$ for $n=1,\ldots,12$, given in
Eq.~(\ref{protocol1sep}). In the latter case, this setup
reduces to the standard QST setup of, e.g.,
Ref.~\cite{James01}. The efficiency of the setup can be
improved if these two polarizers are replaced by PBSs and two
extra detectors are placed at the second outputs of the PBSs,
as in Setup~1. Anyway, this setup is formally simpler but
practically more challenging than Setup~1 because of the use
of the optical CNOT gate. } \label{fig2}
\end{figure}

%------------------------------------------------------------------
\section{Experimental setups for photonic implementations of Protocol~1}

Here let us describe how to implement the optimal QST to
reconstruct an unknown state $\rho$ of two photons by using
polarization degrees of freedom. Thus, in this section, we
assume that the qubit states $\ket{0}$ and $\ket{1}$
correspond to the horizontal $\ket{H}$ and vertical $\ket{V}$
polarizations, respectively.

The projections of a density matrix $\rho$ onto all the
eigenstates $\ket{\psi_{kl}}$ (with $k=1,\ldots,16$ and $l=1$
or $l=1,2$) of the optimal GPOs can be realized
experimentally using Setup~1, shown in Fig.~1 and described
in its caption. The angles of the half-wave plates (HWPs),
$H_1$ and $H_2$, as well as the quarter-wave plates (QWPs),
$Q_1$ and $Q_2$, are given explicitly in Tables~II and~III.
This setup also includes a removable balanced ($50:50$) beam
splitter (BS).

The basic idea is not to directly project $\rho$ onto
$\ket{\psi_{kl}}$, but first to rotate $\rho$ by the HWPs and
QWPs, and only then to project them onto some chosen (and fixed)
states, e.g., either onto $\ket{00}\equiv\ket{HH}$ if the BS is
removed or the singlet state $\ket{\Psi^{-}}$ if the BS is
inserted in Setup~1. Then, the mean values of all the optimal
GPOs, $\gamma_k$, can be calculated directly.

The actions of the HWP and QWP can be defined as
\begin{eqnarray}
  H(\theta) =
  \left[\begin{array}{cccc}
c & s \\
s & -c
\end{array}\right]\!\!,\quad
  Q(\theta) = \tfrac1{\sqrt{2}}
  \left[\begin{array}{cccc}
i+c & s \\
s & i-c
\end{array}\right],
\label{A1}
\end{eqnarray}
given in terms of $c=\cos(2\theta)$ and $s=\sin(2\theta)$. Note
that the operation inverse to the QWP is simply given by
$Q^\dagger(\theta)=-Q(\theta+\pi/2)$. In special cases, $H(0)$,
$H(\pi/8)$, and $H(\pi/4)$ correspond to the phase flip, Hadamard,
and bit flip (NOT) gates, respectively, while $Q(0)$ implements
the phase gate $S$ up to an irrelevant global phase $\phi=-\pi/4$.
Moreover, the circularly-polarized states can be generated from
the horizontally-polarized state as $\ket{R}=-i Q(\pi/4)\ket{H}$
and $\ket{L}=-i Q(-\pi/4)\ket{H}$.

Thus, all the separable eigenstates $\ket{\psi_{kl}}$ of the
optimal GPOs can be transformed into a fixed separable state
$\ket{\psi_{\rm fixed}}$, say equal to $\ket{00}$, by the local
operations implemented by the HWPs and QWPs with the angles
specified in Table~II as follows:
\begin{equation}
  \ket{\psi_{\rm fixed}} =U_{kl}\ket{\psi_{kl}},\; {\rm with}\; U_{kl}={\rm e}^{i\phi_{kl}}
  Q_1^{kl}H_1^{kl} \otimes
  Q_2^{kl}H_2^{kl}
\label{A2}
\end{equation}
for $k=1,\ldots,12$ and $l=1,2. $ Here we have used the
compact notation: $H_{1}^{kl}\equiv
H_{1}(\theta=\theta_{H1}^{kl})$ and $Q_{1}^{kl}\equiv
Q_{1}(\theta=\theta_{Q1}^{kl})$, etc. Moreover, $\phi_{kl}$
are irrelevant global phases.

In order to project a given density matrix $\rho$ onto eight
maximally-entangled eigenstates $\ket{\psi_{kl}}$ of
$\gamma_k$ (for $k=13,\ldots,16$), one can rotate $\rho$ by
the HWPs and QWPs in eight different ways by the angles
specified in, e.g., Table~III. Then we can project all of
them onto the same maximally-entangled state and to perform
its measurement. This Bell measurement can be implemented
efficiently using the central BS in Fig.~1~\cite{KokBook}.

In Table~III, we have assumed  that all the eight global
projections are locally rotated into the singlet state
$\ket{\Psi^{-}}$, which is invariant under the balanced-BS
transformation, according to Eq.~(\ref{B1}). The successful
singlet-state measurement is heralded by the coincidence counts in
either pair of the detectors: $D_{1H}$ and $D_{2V}$ or $D_{1V}$
and $D_{2H}$. Alternatively, one can rotate these global
projections onto the triplet state $\ket{\Psi^{+}}$. This state
after the BS transformation, according to Eq.~(\ref{B2}), is
heralded by the coincidence counts in either pair of the
detectors: $D_{1H}$ and $D_{1V}$ or $D_{2H}$ and $D_{2V}$. In
contrast to these two cases, the projections onto the other six
entangled states, after the BS transformation, cannot be uniquely
distinguished from other orthogonal states in this device, as
explicitly given in Eqs.~(\ref{B3})--(\ref{B5}).

As explained in the caption of Fig.~1, the desired projections
occur probabilistically and they are heralded by proper
coincidence counts. Then, the mean values of the GPOs,
$b_k=\tr(\rho\gamma_k)$, can directly be obtained from the
measured probabilities
\begin{equation}
\tr(\rho\gamma_k)=\sum_l\lambda_{kl}\bra{\psi_{kl}}\rho
\ket{\psi_{kl}}=\sum_l\lambda_{kl}\bra{\psi_{\rm
fixed}}\rho_{kl}\ket{\psi_{\rm fixed}},
\end{equation}
where $\rho_{kl}=U_{kl}\rho U^\dagger_{kl}$ is the rotated density
matrix $\rho$, and $\lambda_{kl}$ (with $l=1$ or $l=1,2$) are
eigenvalues of $\gamma_k$, which can readily be deduced from
Eqs.~(\ref{protocol1sep}), (\ref{protocol1ent}),
and~(\ref{pauli}). Thus, the complete Protocol~1 can be applied.

It is worth noting that the standard setup for photonic QST of,
e.g., James \etal~\cite{James01} can be used to measure all the 20
local projectors $\ket{\psi_{kl}}$ (listed in Table~II). This
would correspond to using Setup~1 without the BS.

Figure~2 shows another setup, which is based on the CNOT gate, or
equivalently the controlled-sign (CS) or iSWAP gates. These gates
can be used for disentangling all the eight maximally-entangled
eigenstates $\ket{\psi_{kl}}$ (for $l=1,2$) into separable eigenstates
$\ket{\psi_{k'l}}$, i.e.,
\begin{eqnarray}
  \ket{\psi_{k'l}} &=& U_{\rm CNOT}\ket{\psi_{kl}}
\label{A3}
\end{eqnarray}
for $k=13,14,15,16$ and $k'=11,12,7,8$, respectively. It is
seen that, contrary to Setup~1, all the rotated states in
this setup are projected onto the same separable state
$\ket{00}$. Thus, this Setup~2 looks formally simpler than
Setup~1. Unfortunately, it is more complicated to be realized
practically because of the use of the optical CNOT gate.
Ref.~\cite{Bartkowiak10} lists more than a dozen
linear-optical implementations of the \emph{nondestructive}
CNOT and/or CS gates including the first proposals of
linear-optical entangling gates~\cite{Knill01,Koashi01}. The
main problem is that these methods usually require some extra
resources including ancillae (separable or entangled),
feedforward, or extra conventional or even single-photon
detectors. Then the complete Setup~2 becomes more complicated
than Setup~1.  Anyway, Setup~2 shows how the optimal QST can,
in principle, be realized also in other systems, where the
CNOT gate can be implemented much more efficiently; for
example, in nuclear-spin devices using NMR spectroscopy
techniques (see, e.g.~\cite{NielsenBook,Hirayama06} and
references therein). Of course, then the HWPs and QWPs have
to be replaced by other feasible local unitary gates, and
photocounts will be replaced by the corresponding NMR
spectra.

An experimental realization of the proposed optimal QST will
be presented elsewhere~\cite{Bart15}. Our results will be
supported there by a numerical simulation of this experiment
assuming realistic single-photon sources including the
generation of the vacuum and multiphoton states. Moreover, we
can  include the effect of imperfect detectors with finite
efficiency, dark counts and their limited (i.e., binary)
resolution. Such a numerical study can be based on
positive-operator-valued measures (POVMs), as applied by us
in, e.g., Ref.~\cite{Ozdemir01} for a related linear-optical
system.

%------------------------------------------------------------------
\section{Optimal tomography of qudit or multiqubit systems}

Finally, we specify the operators, which should be measured
to perform the optimal QST of the density matrix for a
$d$-level qudit (with $d=2,3,\ldots$) or $N$ qubits (where
$d=2^N$ and $N=1,2,\ldots$). Again, the optimality of a QST
method refers to its maximal robustness against errors, as
described by the minimal value of a condition number.

There are $d^2$ unknown real elements of a state $\rho$ of a
$d$-level system, if the normalization of $\rho$ has to  be
determined experimentally. Thus, we need the same number of
optimal GPOs.

Let us analyze the explicit form, as given in
Eq.~(\ref{protocol1}), of the GPOs for the optimal two-qubit
tomography. The symmetry of these two-qubit operators clearly
shows how the optimal tomography can be generalized for
multiqubit or multilevel states. Thus, one can deduce that
the GPOs (denoted as $\gamma^{[d]}_{n}$) for the optimal QST
of a ($d\times d$)-dimensional density matrix can be given as
\begin{equation}
  \{\gamma^{[d]}_{n}; n=1,\ldots,d^2 \}
  = \{X^{[d]}_{k,k}, X^{[d]}_{k<l}, Y^{[d]}_{k<l}; k,l=0,\ldots,d-1
  \},
\label{GPO_N1a}
\end{equation}
where
\begin{eqnarray}
  X^{[d]}_{k,l} &=& \tfrac12 (\ket{k}\bra{l}+\ket{l}\bra{k}),\\
  Y^{[d]}_{k,l} &=& \tfrac{1}{2} (-i\ket{k}\bra{l}+i\ket{l}\bra{k}).
\label{GPO_N1b}
\end{eqnarray}
The measurement of the each GPO, $\gamma^{[d]}_{n}$, corresponds
to a direct measurement of either the real ($X^{[d]}_{i,j}$) or
imaginary ($Y^{[d]}_{i,j}$) part of the element $\rho_{ij}$ of a
given density matrix $\rho$. By this construction, it is seen that
the QST based on the measurement of $\gamma^{[d]}_{n}$ is optimal
as described by the condition number $\kappa(A)=1$.

For a comparison, let us analyze another simple approach to
QST of $N$ qubits, which is based on the local measurements of
the tensor products of the standard single-qubit Pauli
operators, i.e.,
\begin{eqnarray}
 \Gamma^{[d]}_{n} &=& \sigma_{n_1}\otimes \sigma_{n_2}
 \otimes \ldots \otimes\sigma_{n_N},
\label{GPO_N2}
\end{eqnarray}
where $n=\{n_1,n_2,\ldots,n_N\}$ and $n_i=0,1,2,3.$ This
multiindex $n$ can be considered  a single number written in
the ternary numeral system, i.e., $n = 1+\sum_{i=1}^N
4^{N-i}n_i$. The condition number for this method is given by
$\kappa(C)=\kappa^2(A)=2$. Thus, it is seen that this simple
approach, based on local measurements, is less robust against
errors in comparison to our approach, based on both local and
global measurements. Moreover, the latter method can be
applied for the QST of a single $d$-level system only if
$d=2^N$. Note that our GPO-based method can be used for
qudits with any number of levels even if $d\neq 2^N$.

%------------------------------------------------------------------
\subsection{Single-qubit tomography}

For clarity and completeness of our presentation, let us
analyze the simplest case of the reconstruction of a single-qubit
density matrix,
\begin{eqnarray}
  \rho &=& \left[
\begin{array}{cc}
 x_1 & x_2+i x_3 \\
 x_2-i x_3 & x_4 \\
\end{array}
\right]. \label{qubit1}
\end{eqnarray}
By applying the optimal projectors $\gamma^{[1]}_{n}$, given by
\begin{eqnarray}
\left[
\begin{array}{cc}
 1 & 0 \\
 0 & 0 \\
\end{array}
\right],\left[
\begin{array}{cc}
 0 & 0 \\
 0 & 1 \\
\end{array}
\right],\frac12\left[
\begin{array}{cc}
 0 & 1 \\
 1 & 0 \\
\end{array}
\right], \frac12\left[
\begin{array}{cc}
 0 & -i \\
 i & 0 \\
\end{array}
\right], \label{GPO1}
\end{eqnarray}
one  finds that the protocol is optimal, as described by
\begin{eqnarray}
  A &=&
\left[
\begin{array}{cccc}
 1 & 0 & 0 & 0 \\
 0 & 0 & 0 & 1 \\
 0 & 1 & 0 & 0 \\
 0 & 0 & -1 & 0 \\
\end{array}
\right], \label{A4opt}
\end{eqnarray}
with the condition number $\kappa(A)=1$.

For a comparison, let us analyze the standard approach of
finding four unknowns $x_n$ (for $n=1,\ldots,4$) by applying
the Pauli operators together with the identity operator,
$\{\sigma_1,\sigma_2,\sigma_3,I\}$, which is  a special case
of Eq.~(\ref{GPO_N2}). A simple calculation shows that
\begin{eqnarray}
  A &=&
\left[
\begin{array}{cccc}
 0 & 2 & 0 & 0 \\
 0 & 0 & -2 & 0 \\
 1 & 0 & 0 & -1 \\
 1 & 0 & 0 & 1 \\
\end{array}
\right], \label{A4Pauli}
\end{eqnarray}
which leads to the condition number $\kappa(C)=\kappa^2(A)=2$. Thus,
this approach is not optimal.

Note that this standard approach is optimal for finding only three
unknowns $x_n$ (for $n=1,2,3$), as given by Eq.~(\ref{qubit1}),
where $x_4$ is determined from the normalization condition, as
$x_4=1-x_1$, and the projectors include only the Pauli operators
without the identity operator. In this case, one finds that
\begin{eqnarray}
  A &=&
2\left[
\begin{array}{ccc}
 0 & 1 & 0 \\
 0 & 0 & -1 \\
 1 & 0 & 0 \\
\end{array}
\right], \label{A3Pauli}
\end{eqnarray}
for which the condition number is $\kappa(A)=1$, as desired.
In this case, the corresponding observation vector $b$ should
be displaced as, $b\rightarrow b+[0,0,1]$.

In the concluding section, we address the problem of reducing
the number of variables by applying the normalization
condition and its effect on the error robustness of two-qubit
tomography.

%------------------------------------------------------------------
\section{Discussion and conclusions}

The main problem studied here was to find a QST method which
is the most robust against errors, as described by the
condition numbers defined via the spectral norm (i.e., the
two-norm condition number).

If QST is directly based on solving a linear-system problem, given
in Eq.~(\ref{N1}), then the condition number $\kappa(A)$ is a good
measure of the QST robustness against errors in the observation
vector $b$. Indeed, according to the Gastinel-Kahan
theorem~\cite{Kahan}, the condition number $\kappa(A)$ has a clear
geometric meaning as the reciprocal of a relative distance of a
nonsingular matrix $A$ to the set of singular matrices. The
smaller is the condition number the more robust is the QST method,
and the optimal method is described by $\kappa(A)=1$.

The main advantage of using condition numbers to describe the
error robustness of QST methods might be that the condition
numbers determine lower and upper bounds, as given by
Eqs.~(\ref{Atkinson2}) and~(\ref{Atkinson3}), on the errors
in a reconstructed state $\rho$. This estimation of error
robustness depends on the errors in the measured data,
although the real sources of the errors are irrelevant. For
example, they can be related to imperfect detectors,
realistic photon sources, lossy and unbalanced linear-optical
elements (including beam splitters and wave plates), etc.

We found such a QST method (referred to as Protocol~1), based
on the measurement of the generalized Pauli operators,
defined in Eqs.~(\ref{protocol1sep})--(\ref{protocol1ent}).
Protocol~1 corresponds to measuring one by one all of the
real and imaginary elements of an unknown two-qubit density
matrix. This approach results in the condition number
$\kappa(A)=1$. Thus, Protocol~1 can be considered as the most
robust against errors, which can occur in the observation
vector $b$. Moreover, we described two experimentally
feasible setups of photonic implementations of Protocol~1.

In Table~I, we compared this error robustness of Protocol~1
with six other QST methods: Protocol~2 is based on the
measurements of all the 16 tensor products of the standard
Pauli operators, $\sigma_i\otimes\sigma_j$. Protocol~3 is the
well-known QST method of James \etal~\cite{James01} based on
16 projectors given explicitly in Appendix~A. Protocol~4,
often referred to as standard-separable QST, is based on the
measurements of all the 36 eigenstates of the operators
$\sigma_i\otimes\sigma_j$ used in Protocol~2. Protocol~5 is
based on the projections onto MUB according to the original
idea of Wootters and Fields~\cite{Wootters89}, later studied
in, e.g., Refs.~\cite{Bandyopadhyay01,Aravind03,DAriano05},
and experimentally applied by Adamson and
Steinberg~\cite{Adamson10}. Protocols~6 and~7 are based on
the measurements of the Gell-Mann GPOs for the special
unitary group SU(4) (see, e.g., Ref.~\cite{Kimura03}) and the
Patera-Zassenhaus GPOs for the general linear group
GL(4,C)~\cite{Patera88}, respectively. The projectors for all
these seven protocols are defined explicitly in Appendix~A.

It is worth noting that the projections of $\rho$ onto all the
eigenstates of tensor products of the standard Pauli operators
$\sigma_i\otimes\sigma_j$ in Protocol~4 also enable the
application of Protocol~2 based on the measurements of
$\sigma_i\otimes\sigma_j$. Namely, the set of 16 equations in
Protocol~2 can be obtained by proper linear combinations of the 36
equations in Protocol~4, according to the eigenstate expansions of
the Pauli operators, given in Eq.~(\ref{pauli}). Thus, the error
robustness of QST can be improved 4.5 times if described by the
conditions numbers $\kappa(C)=\kappa^2(A)$ (see Table~I). This
approach, based on pure-state projections of an unknown state
$\rho$, can also be applied to measure the GPOs in Protocols~1, 7,
and~8.

Protocols~2--5 are based solely on local rotations and local
measurements. However, all the other protocols require both
single-qubit and nonlocal two-qubit projections. These
nonlocal projections of an unknown state onto a given Bell
state can be realized effectively in the standard Bell
analyzers~\cite{NielsenBook,KokBook} as applied in Setups~1
and 2 presented in Figs.~1 and 2. Note that
Ref.~\cite{Adamson10} describes not only a proposal to use
MUB for tomography (referred to here as Protocol~5), but also
reports an experimental photonic implementation, which
includes nonlocal projections of a given state onto Bell-like
states, which corresponds to Setup~1 in Fig.~1.

Protocols~1 and~6 are apparently similar. Indeed, all twelve
nondiagonal Gell-Mann GPOs are the same as the optimal GPOs,
i.e., $\Gamma^{(6)}_{n} = \gamma_{n}$, for $n=5,\ldots,16$.
However, the other four diagonal Gell-Mann GPOs are different
from $\gamma_{n}$. Note that the Gell-Mann GPOs, like the
standard Pauli matrices, are Hermitian, traceless, and
orthogonal in the Hilbert-Schmidt inner product. While the
optimal GPOs $\gamma_n$ are Hermitian and orthogonal, but the
diagonal ones are not traceless. This difference implies that
the error robustness of Protocol~6 is twice worse than that
of Protocol~1, and this is the same as in Protocol~2 being
solely based on local measurements. Thus, the
nonlocal-projections in Protocol~6 do not offer any advantage
(in terms of the condition number) over the local ones in
Protocol~2.

Surprisingly, the MUB-based Protocol~5 is nonoptimal
concerning the error robustness, measured by the condition
numbers $\kappa(C)=\kappa^2(A)$, which is five times worse
than the optimal Protocol~1 and $\frac{5}{2}$ times worse
than Protocol~2 based solely on local measurements.

It is quite counterintuitive that the projections onto MUB of
Refs.~\cite{Bandyopadhyay01,Aravind03,DAriano05,Adamson10}
are the nonoptimal choices of measurements in terms of the
lowest condition number. Nevertheless, nonoptimality of MUB
was also observed in other contexts, e.g., in the detection
of the Einstein-Podolski-Rosen steering, where random
measurements are in some cases better than MUB (maximally
noncommuting observables)~\cite{Skrzypczyk14}. It should be
also noted that the MUB-based Protocol~5 is the most robust
against errors among the QST protocols (listed in Table~I),
which are based solely on pure-state projections.

It should be stressed that we discussed the reconstruction of
16 real elements of an unknown two-qubit density matrix
$\rho$. One could argue that only 15 elements are unknown,
since the 16th element can be calculated from the
normalization condition $\tr\rho=1$. However, in the
experiments with imperfect detection efficiency (like typical
photon counting), this normalization has to be determined in
a separate measurement (corresponding to a separate
equation). Thus, one has to determine all the 16 unknown
elements in such experiments. In particular, the MUB-based
reconstruction of only 15 elements, results in the case of
perfect error robustness as described by the condition number
$\kappa(A)=1$. However, by including the 16th unknown element
of $\rho$ (say $\rho_{44}$), the error robustness of this MUB
approach is five times worse. As explained above, we prefer
to reconstruct all the 16 elements for operational reasons.
To clarify this point let us give a typical example of a
two-photon state, where $\rho$ is unnormalized. By using the
parametrization of $\rho$, given in Eq.~(\ref{rho}), one can
say that $\rho$ corresponds to $\eta\rho'$, where $\rho'$ is
the normalized density operator of the measured two-photon
state and $\eta$ is the unknown efficiency for detecting the
two photons. By knowing all the diagonal terms of $\rho$, one
can directly determine $\eta$ as $\tr\rho$.

We also generalized our approach by defining observables for
the optimal reconstruction of the unknown state of an
arbitrary number of qubits or arbitrary-level qudits. This
method is the most robust against errors, since $\kappa(A)=1$
for any dimension of the state. For a comparison, we analyzed
a simple approach to QST of a multiqubit system based on the
local measurements of the tensor products of the standard
single-qubit Pauli operators. The latter approach is not
optimal, as described by the condition number
$\kappa(A)=\sqrt{2}$.

Finally, we express our hope that the proposed tomographic
protocol, which is optimally robust against errors and can be
easily implemented by using, e.g., linear optics, can become
a useful tool for quantum engineering and quantum information
processing.

%------------------------------------------------------------------
\begin{acknowledgments}
The authors thank Anton\'in \v{C}ernoch, Daoyi Dong, Karel
Lemr, \c{S}ahin \"Ozdemir, Bo Qi, and Jan Soubusta for
discussions. A.M. is supported by the Polish National Science
Centre under Grants DEC-2011/03/B/ST2/01903 and
DEC-2011/02/A/ST2/00305. K.B. acknowledges the support by the
Polish National Science Centre (Grant No.
DEC-2013/11/D/ST2/02638) and by the Foundation for Polish
Science (START Programme). K.B. and J.P. are supported by the
project No. LO1305 of the Ministry of Education, Youth and
Sports of the Czech Republic. N.I. is supported by JSPS
Grant-in-Aid for Scientific Research(A) 25247068. F.N. is
partially supported by the RIKEN iTHES Project, MURI Center
for Dynamic Magneto-Optics, and a Grant-in-Aid for Scientific
Research (S).
\end{acknowledgments}

%%%%%%%%%%%%%%%%%%%%%%%%%%%%%%%%%%%%%%%%%%%%%%%%%%%%%%%%%%%%%%%%%%%%%%%%%%%%%%%%%%%
\appendix

\section{Projectors for quantum-state-tomography protocols}

For the benefit of the reader, we explicitly show here the QST
projectors and other details for the QST protocols discussed in
Table~I.

\section*{Protocol 1 with optimal generalized Pauli operators}

The two-qubit optimal GPOs, given by Eqs.~(\ref{protocol1sep}) and
(\ref{protocol1ent}), have the following symmetrical forms in the
standard computational basis:
\begin{eqnarray}
\gamma_{1}&=&{\rm diag}([1 ,0, 0, 0]),
\quad\;\; \gamma_{2}={\rm diag}([0 ,1, 0, 0]),
 \nonumber \\
\gamma_{3}&=&{\rm diag}([0,0, 1, 0]),
\quad\;\; \gamma_{4}={\rm diag}([0 ,0,0, 1]),
 \nonumber  \\
\gamma_{5}&=&\frac{1}{2}\left[\begin{array}{cccc}
0 & 1 & 0 & 0\\
1 & 0 & 0 & 0\\
0 & 0 & 0 & 0\\
0 & 0 & 0 & 0
\end{array}\right],\quad\gamma_{6}=\frac{1}{2}\left[\begin{array}{cccc}
0 & -i & 0 & 0\\
i & 0 & 0 & 0\\
0 & 0 & 0 & 0\\
0 & 0 & 0 & 0
\end{array}\right],
 \nonumber  \\
\gamma_{7}&=&\frac{1}{2}\left[\begin{array}{cccc}
0 & 0 & 1 & 0\\
0 & 0 & 0 & 0\\
1 & 0 & 0 & 0\\
0 & 0 & 0 & 0
\end{array}\right],\quad\gamma_{8}=\frac{1}{2}\left[\begin{array}{cccc}
0 & 0 & -i & 0\\
0 & 0 & 0 & 0\\
i & 0 & 0 & 0\\
0 & 0 & 0 & 0
\end{array}\right],
 \nonumber  \\
\gamma_{9}&=&\frac{1}{2}\left[\begin{array}{cccc}
0 & 0 & 0 & 0\\
0 & 0 & 0 & 0\\
0 & 0 & 0 & 1\\
0 & 0 & 1 & 0
\end{array}\right],\quad\gamma_{10}=\frac{1}{2}\left[\begin{array}{cccc}
0 & 0 & 0 & 0\\
0 & 0 & 0 & 0\\
0 & 0 & 0 & -i\\
0 & 0 & i & 0
\end{array}\right],
 \nonumber  \\
\gamma_{11}&=&\frac{1}{2}\left[\begin{array}{cccc}
0 & 0 & 0 & 0\\
0 & 0 & 0 & 1\\
0 & 0 & 0 & 0\\
0 & 1 & 0 & 0
\end{array}\right],\quad\gamma_{12}=\frac{1}{2}\left[\begin{array}{cccc}
0 & 0 & 0 & 0\\
0 & 0 & 0 & -i\\
0 & 0 & 0 & 0\\
0 & i & 0 & 0
\end{array}\right],
 \nonumber  \\
\gamma_{13}&=&\frac{1}{2}\left[\begin{array}{cccc}
0 & 0 & 0 & 0\\
0 & 0 & 1 & 0\\
0 & 1 & 0 & 0\\
0 & 0 & 0 & 0
\end{array}\right],\quad\gamma_{14}=\frac{1}{2}\left[\begin{array}{cccc}
0 & 0 & 0 & 0\\
0 & 0 & -i & 0\\
0 & i & 0 & 0\\
0 & 0 & 0 & 0
\end{array}\right],
 \nonumber  \\
\gamma_{15}&=&\frac{1}{2}\left[\begin{array}{cccc}
0 & 0 & 0 & 1\\
0 & 0 & 0 & 0\\
0 & 0 & 0 & 0\\
1 & 0 & 0 & 0
\end{array}\right],\quad\gamma_{16}=\frac{1}{2}\left[\begin{array}{cccc}
0 & 0 & 0 & -i\\
0 & 0 & 0 & 0\\
0 & 0 & 0 & 0\\
i & 0 & 0 & 0
\end{array}\right].
 \nonumber \\
\label{protocol1}
\end{eqnarray}
For this protocol, the condition numbers are minimal,
$\kappa(C)=\kappa(A)=1$.

\section*{Protocol 2 with standard Pauli operators}

Protocol 2 for two-qubit QST is based on measuring all the tensor
products of the single-qubit Pauli operators (see, e.g.,
Ref.~\cite{James01} and references therein):
\begin{eqnarray}
  \Gamma^{(2)}_{4i+j+1} &=& \sigma_i \otimes \sigma_j
\label{protocol2}
\end{eqnarray}
for $i,j=0,\ldots,3$, where $\sigma_0=I$ is the identity
operator. This is a natural generalization of the
single-qubit QST. For this protocol, the condition numbers
are $\kappa(C)=\kappa^2(A)=2$.

\section*{Protocol 3 of James et al.}

Protocol 3 for QST is based on the following projections
$\{\ket{\psi^{(3)}_n} \}$, which were applied in the QST
experiment performed by James et al.~\cite{James01}:
\begin{eqnarray}
\{ \ket{\psi^{(3)}_n} \} &=& \big\{
 \ket{00}, \ket{01}, \ket{0+}, \ket{0L},
\nonumber \\ &&
 \ket{10}, \ket{11}, \ket{1+}, \ket{1L},
\nonumber \\ &&
 \ket{R0}, \ket{R1}, \ket{R+}, \ket{RL},
\nonumber \\ &&
 \ket{+0}, \ket{+1}, \ket{++},  \ket{+R}
 \big\}.
\label{protocol3}
\end{eqnarray}
The resulting condition numbers are the largest among the studied
protocols, as $\kappa(C)=\kappa^2(A)\approx 60.1$.

\section*{Protocol 4 with  Pauli operator eigenstates}

Protocol 4 is probably the most popular experimental
two-qubit QST method and is referred to as standard-separable
QST. It is based on the projections onto all of the 36 tensor
products of the eigenstates of the standard single-qubit
Pauli operators~\cite{Altepeter05} (see also~\cite{Burgh08}):
\begin{eqnarray}
\{ \ket{\psi^{(4)}_n} \} &=& \big\{
\ket{00}, \ket{01}, \ket{10}, \ket{11},\ket{\pm+},
\ket{\pm-},
\nonumber \\ &&
\ket{0\pm},\ket{\pm\!0}, \ket{1\pm},\ket{\pm\!1},
\nonumber \\ &&
\ket{0R}, \ket{R0}, \ket{1R},\ket{R1},
%\nonumber \\ &&
\ket{0L},\ket{L0},  \ket{1L},\ket{L1},
\nonumber \\ &&
\ket{R\pm},\ket{\pm\!R},   \ket{L\pm},
\ket{\pm\!L},
\nonumber \\ &&
\ket{RR}, \ket{RL}, \ket{LR}, \ket{LL}
\big\}.
\label{protocol4}
\end{eqnarray}
The corresponding condition numbers are $\kappa(C)=\kappa^2(A)=9$.

\section*{Protocol 5 based on mutually unbiased bases}

Protocol~5 is based on the five MUB of Adamson and
Steinberg~\cite{Adamson10}:
\begin{eqnarray}
\{
\ket{\psi^{(5)}_n} \}=\{
\ket{\psi^{A}_n},
\ket{\psi^{B}_n},
\ket{\psi^{C}_n},
\ket{\psi^{D}_n},
\ket{\psi^{E}_n} \},
\label{protocol5}
\end{eqnarray}
where
\begin{eqnarray}
  \{ \ket{\psi^{A}_n} \} &=& \big\{ \ket{00}, \ket{01}, \ket{10},
  \ket{11}\big\},
 \nonumber \\
  \{ \ket{\psi^{B}_n} \} &=& \big\{ \ket{R\pm}, \ket{L\pm}\big\},
  \nonumber \\
  \{ \ket{\psi^{C}_n} \} &=& \big\{\ket{\pm\! R},  \ket{\pm\! L}\big\},
 \nonumber \\
\{ \ket{\psi^{D}_n} \} &=& \big\{ \fra{1}{\sqrt{2}}( \ket{R0}\pm i \ket{L1}), \fra{1}{\sqrt{2}}( \ket{R1}\pm i \ket{L0})\big\},
 \nonumber \\
\{ \ket{\psi^{E}_n} \} &=& \big\{ \fra{1}{\sqrt{2}}( \ket{RR}\pm i \ket{LL}), \fra{1}{\sqrt{2}}( \ket{RL}\pm i \ket{LR})
  \big\}.\quad\quad
\label{protocol5a}
\end{eqnarray}
This protocol was applied in the QST experiment of
Ref.~\cite{Adamson10}. The twenty states of the MUB include twelve
separable  and eight Bell-like states. The latter are simply
related to the standard Bell states by local operations as
follows:
\begin{eqnarray}
  \ket{\psi^{D}_{1,2}} &=& \frac{1}{\sqrt{2}}( \ket{R0}\pm i \ket{L1}) \cong (SHS\otimes {\sigma_2}) \ket{\Phi^{\pm}} ,
  \nonumber \\
  \ket{\psi^{D}_{3,4}} &=& \frac{1}{\sqrt{2}}( \ket{R1}\pm i \ket{L0}) \cong (SHS\otimes {\sigma_2}) \ket{\Psi^{\pm}},
 \nonumber \\
  \ket{\psi^{E}_{1,2}} &=&  \frac{1}{\sqrt{2}}( \ket{RR}\pm i \ket{LL}) \cong (SHS\otimes SH) \ket{\Phi^{{\pm}}},
  \nonumber \\
  \ket{\psi^{E}_{3,4}} &=& \frac{1}{\sqrt{2}}( \ket{RL}\pm i \ket{LR}) \cong (SHS\otimes SH) \ket{\Psi^{{\pm}}},\quad\quad
\label{epr}
\end{eqnarray}
where $S$ and $H$ are  the phase and Hadamard gates, respectively,
and the sign $\cong$ indicates that the corresponding expressions
are equal up to irrelevant global phase factors.

Another MUB for two qubits was studied by Bandyopadhyay
\etal~\cite{Bandyopadhyay01} and
others~\cite{Aravind03,DAriano05}. Here we rewrite this MUB
explicitly in terms of the twelve separable and eight
Bell-like states, analogously with Eq.~(\ref{protocol5a}). We
have
\begin{eqnarray}
   \{ \ket{\psi^{B}_n} \} &=& \big\{ \ket{\pm+},\ket{\pm-}\big\},
  \nonumber \\
  \{ \ket{\psi^{C}_n} \} &=& \big\{\ket{RR}, \ket{RL},\ket{LR},\ket{LL}\big\},
 \nonumber \\
\{ \ket{\psi^{D}_n} \} &=& \big\{ U_{1}\ket{\Phi^{\pm}},
 U_{1}\ket{\Psi^{\pm}} \big\}
  \nonumber \\
   &\cong&\big\{\tfrac{1}{\sqrt{2}}( \ket{L0}\pm \ket{R1}),
                \tfrac{1}{\sqrt{2}}( \ket{R0}\pm \ket{L1}) \big\},
 \nonumber \\
\{ \ket{\psi^{E}_n} \} &=& \big\{ U_{2}\ket{\Phi^{\pm}},
 U_{2}\ket{\Psi^{\pm}}
  \big\}
  \nonumber \\
   &=&\big\{\tfrac{1}{\sqrt{2}}( \ket{0L}\pm \ket{1R}),
                \tfrac{1}{\sqrt{2}}( \ket{0R}\pm \ket{1L}) \big\},
\label{protocol5b}
\end{eqnarray}
and the basis $\{ \ket{\psi^{A}_n} \}$ is the same as in
Eq.~(\ref{protocol5a}). Moreover, $U_{1}=SH\otimes I$ and
$U_{2}=I\otimes SH$.

It is easy to confirm that both Eqs.~(\ref{protocol5a}) and
(\ref{protocol5b}) represent MUB because it holds
\begin{eqnarray}
  |\langle{\psi^X_m}\ket{\psi^Y_n}| = \frac12
\label{Mubdef}
\end{eqnarray}
for any $m,n\in\{1,\ldots,4\}$ and $X\neq
Y\in\{A,\ldots,E\}$.

The condition numbers $\kappa(C)=\kappa^2(A)$ are equal to 5
for both of these MUB, given by Eqs.~(\ref{protocol5a})
and~(\ref{protocol5b}).

%------------------------------------------------------------------
\section*{Protocol~6 with Gell-Mann generalized Pauli operators}

Protocol~6 is based on the Gell-Mann GPOs for the special unitary
group SU(4). These GPOs can be given in the standard computational
basis (see, e.g., Ref.~\cite{Kimura03}) as
\begin{eqnarray}
\Gamma^{(6)}_{1} & = & \frac{1}{2}I_4,\nonumber \\
\Gamma^{(6)}_{2} & = & \frac{1}{2} \;{\rm diag} ([1,-1,0,0]),\nonumber \\
\Gamma^{(6)}_{3} & = & \frac{1}{2\sqrt{3}}\; {\rm diag} ([1,1,-2,0]),\nonumber \\
\Gamma^{(6)}_{4} & = & \frac{1}{2\sqrt{6}}\; {\rm diag} ([1,1,1,-3]),\nonumber \\
\Gamma^{(6)}_{n} & = & \gamma_{n}\quad {\rm for}\quad
n=5,\ldots,16, \label{protocol7}
\end{eqnarray}
where $\gamma_{n}$ are our GPOs, defined in Eq.~(\ref{protocol2}),
and $I_4$ is the four-dimensional identity operator. The
corresponding condition numbers are $\kappa(C)=\kappa^2(A)= 2$.

\section*{Protocol~7 with Patera-Zassenhaus generalized Pauli operators}

Protocol~7 is based on the Patera-Zassenhaus GPOs for the general
linear group GL(4,C)~\cite{Patera88}:
\begin{eqnarray}
\Gamma^{(7)}_{1} & = & D,\quad \quad  \Gamma^{(7)}_{2} = D^2,\quad \quad \Gamma^{(7)}_{3} = D^3,\nonumber \\
\Gamma^{(7)}_{4} & = & B, \quad \quad  \Gamma^{(7)}_{5} =
B^2,\quad \quad\;
\Gamma^{(7)}_{6} = B^3,\nonumber \\
\Gamma^{(7)}_{7} & = & BD,\quad \; \Gamma^{(7)}_{8} = BD^2,\quad\; \Gamma^{(7)}_{9} = BD^3,\nonumber \\
\Gamma^{(7)}_{10} & = & B^2D, \quad \Gamma^{(7)}_{11} = B^2D^2,
\quad \Gamma^{(7)}_{12} = B^2D^3,\nonumber \\
\Gamma^{(7)}_{13} & = & B^3D, \quad \Gamma^{(7)}_{14} = B^3D^2,
\quad \Gamma^{(7)}_{15} = B^3D^3,\quad \label{protocol8}
\end{eqnarray}
and $\Gamma^{(7)}_{16} = I_4$, where
\begin{eqnarray}
D&=&\exp(i \pi/4)\;{\rm diag}
 ([1,i,-1,-i]),
\nonumber \\
  B&=&\left[
\begin{array}{cccc}
 0 & 1 & 0 & 0 \\
 0 & 0 & 1 & 0 \\
 0 & 0 & 0 & 1 \\
 -1 & 0 & 0 & 0 \\
\end{array}
\right].
  \label{Patera2}
\end{eqnarray}
The corresponding condition numbers for this protocol are the same
as for Protocols~2 and~7, i.e., $\kappa(C)=\kappa^2(A)=2$.

%------------------------------------------------------------------
\section{Beam-splitter transformation of entangled projectors}

Here we show explicitly the transformation via a $50:50$ beam
splitter of the Bell and Bell-like states, which are the
entangled projectors, i.e., the eight maximally-entangled
eigenstates of the optimal GPOs $\gamma_n$ for
$n=13,\ldots,16$.

The $50:50$ beam-splitter transformation $U_{\rm BS}$ can be
implicitly given by the transformation between the input
($a_{1p}$ and $a_{2p}$) and output $(b_{1p}$ and $b_{2p}$)
annihilation operators, e.g.,~\cite{KokBook,Bart13}:
$a_{1p}=(b_{1p}+b_{2p})/\sqrt{2}$ and
$a_{2p}=(b_{1p}-b_{2p})/\sqrt{2}$, for two polarizations
$p=H,V$. Then the entangled projectors are transformed as
follows
\begin{eqnarray}
  U_{\rm BS} \ket{\Psi^{-}} &=& -\ket{\Psi^{-}},
\label{B1}\\
  U_{\rm BS} \ket{\Psi^{+}} &=& \frac{1}{\sqrt{2}}(\ket{HV,\vac}-\ket{\vac,HV}),
\label{B2}
\\
  U_{\rm BS} \ket{\Phi^{\pm}} &=&
  \frac{1}{2}(\ket{2H,\vac}-\ket{\vac,2H})
  \nonumber  \\ && \pm \frac{1}{2}(\ket{2V,\vac}-\ket{\vac,2V}),
\label{B3}
\\
  U_{\rm BS} \ket{\bar\Psi^{\pm}} &=&
  c^{\pm}(\ket{HV,\vac}-\ket{\vac,HV})
  \nonumber\\&&- c^{\mp}(\ket{H,V}-\ket{V,H})
\nonumber\\
  &=& \sqrt{2}c^{\pm}U_{\rm BS}\ket{\Psi^{+}}- \sqrt{2}c^{\mp}\ket{\Psi^{-}},
 \label{B4}\\
  U_{\rm BS} \ket{\bar\Phi^{\pm}} &=&
  \frac{1}{2}(\ket{2H,\vac}-\ket{\vac,2H})
  \nonumber\\&&\pm \frac{i}{2}(\ket{2V,\vac}-\ket{\vac,2V}),
  \label{B5}
\end{eqnarray}
where $\ket{\vac}$ stands for the vacuum, $c^{\pm}=(1\pm
i)/(2\sqrt{2})$,
$\ket{V,H}=b_{1V}^{\dagger}b_{2H}^{\dagger}\ket{\vac,\vac}$, and
$\ket{2V,\vac}=\tfrac1{\sqrt{2}}
b_{1V}^{\dagger2}\ket{\vac,\vac}$, etc.

%------------------------------------------------------------------

\end{document}